\documentclass[12pt]{article}
\usepackage{amsmath}
\usepackage{graphicx,psfrag,epsf}
\usepackage{enumerate}
\usepackage{natbib}
\usepackage{url} 
\usepackage{amsmath,amssymb}
\usepackage[dvips]{color}
\usepackage{mathrsfs}
\usepackage{bm, xcolor}
\usepackage{mathrsfs}
\usepackage{verbatim}
\usepackage{threeparttable}
\usepackage{color}
\usepackage[colorlinks,
linkcolor=blue,
anchorcolor=blue,
citecolor=blue,
breaklinks=true]{hyperref}
\usepackage{xr-hyper}
\usepackage[utf8]{inputenc}
\usepackage{epstopdf}
\usepackage{booktabs}
\usepackage{graphicx}
\usepackage{float}
\usepackage{listings}
\usepackage{amssymb}
\usepackage{amsthm}
\usepackage{multirow}
\usepackage{rotating}
\usepackage{longtable}
\usepackage{enumitem}
\usepackage{threeparttablex} 
\usepackage{booktabs} 
\usepackage{cases}
\usepackage{filecontents}
\usepackage[section]{placeins}
\usepackage{subcaption} 
\usepackage{algorithm}
\usepackage{algorithmic}
\usepackage{caption}
\usepackage{setspace}

\newtheorem{theorem}{Theorem}[section]
\newtheorem{corollary}{Corollary}[section]

\newtheorem{proposition}{Proposition}[section]

\newtheorem{assumption}{Assumption}[section]

\newcommand{\blind}{0}

\numberwithin{equation}{section}
\begin{document}

\def\spacingset#1{\renewcommand{\baselinestretch}%
{#1}\small\normalsize} \spacingset{1}


\if0\blind
{
  \title{\bf Inference of high-dimensional weak instrumental variable regression models without ridge-regularization}

  \author{Jiarong Ding$^1$, Xu Guo$^1$, Yanmei Shi$^{1}$\thanks{
  Corresponding author: Yanmei Shi. Email address: ymshi@mail.bnu.edu.cn. 
}, and Yuxin Wang$^2$
  	\\	{\small \it $^{1}$ School of Statistics, Beijing Normal University, Beijing, China}\\
    {\small \it $^{2}$ School of Economics, Hefei University of Technology, Hefei, China}
  }

  \date{}
  \maketitle
} \fi

\if1\blind
{
  \bigskip
  \bigskip
  \bigskip
  \begin{center}
    {\LARGE\bf Title}
\end{center}
  \medskip
} \fi

\bigskip
\begin{abstract}

Inference of instrumental variable regression models with many weak instruments attracts many attentions recently. 
To extend the classical Anderson-Rubin test to high-dimensional setting, many procedures adopt ridge-regularization. However, we show that it is not necessary to consider ridge-regularization. Actually we propose a new quadratic-type test statistic which does not involve tuning parameters. 
Our quadratic-type test exhibits high power against dense alternatives. While for sparse alternatives, we derive the asymptotic distribution of an existing maximum-type test, enabling the use of less conservative critical values. To achieve strong performance across a wide range of scenarios, we further introduce a combined test procedure that integrates the strengths of both approaches. This combined procedure is powerful without requiring prior knowledge of the underlying sparsity
of the first-stage model. 
Compared to existing methods, our proposed tests are easy to implement, free of tuning parameters, and robust to arbitrarily weak instruments as well as heteroskedastic errors. Simulation studies and empirical applications demonstrate the advantages of our methods over existing approaches.
\end{abstract}

\noindent%
\vfill

\newpage
\spacingset{1.8} 
\section{Introduction} \label{section1}


One of the classic problems in econometrics \citep{wang1998inference, hahn2003weak, andrews2019weak} and epidemiology \citep{glymour2006natural, lawlor2008mendelian, davey2014mendelian} is making inference on the coefficients of structural models. These models typically involve  endogenous variables, which introduce endogeneity biases.  To address this issue, researchers commonly employ instrumental variables (IVs) to achieve identifiability of structural parameters of interest.
However, the effectiveness of IV methods critically depends on both the quantity and strength of the instruments. A central concern in this context is ``weak instruments", i.e., IVs are only weakly correlated with the endogenous regressors. Weak instruments undermine the reliability of conventional  inference procedures, such as  two-stage least squares (2SLS) t-tests, and leads to poor finite sample performance \citep{andrews2019weak}.
As a result, effectively addressing weak identification, especially in high-dimensional settings with numerous weak instruments (so-called ``many weak instruments"), has become increasingly  critical. Motivated by this challenge, this paper aims to develop inference procedures for coefficients of endogenous variables in the presence of many weak IVs.

We focus on inference methods that are robust to weak identification. Existing literature has  demonstrated that the conventional procedures often perform inadequately when instruments are weak \citep{kleibergen2002pivotal,moreira2003conditional,moreira2009tests}. Consequently, alternative testing approaches that are robust to weak IVs have been proposed, among which the Anderson-Rubin (AR) test \citep{anderson1949estimation} is prominent.  However, the finite-sample performance of the AR test deteriorates as the dimensionality of the IVs increases \citep{anatolyev2011specification}.
In addition,  \cite{andrews2007testing} showed that although the AR test maintains correct asymptotic size under many instruments, it requires the condition $K^3/n \rightarrow 0$ as $n \rightarrow \infty$ and reduced-form errors are homoskedastic. Here $K$ is the number of instrumental variables and $n$ is the sample size. To address these issues, a variety of improvements to the AR test have emerged.
For example,
\cite{newey2009generalized} proposed a generalized method of moments (GMM) variant of AR test that accommodates heteroskedasticity, yet imposes the same condition on $K$.
More recently, \cite{bun2020finite} introduced a degrees-of-freedom correction for the centered GMM-AR test under the condition that $ K/n \rightarrow 0$.  Comparable restrictions are also found in \cite{phillips2017structural} under conditional homoskedasticity.
We note that these methods were developed mainly under conditions of ``moderately many IVs", where $K$ is asymptotically negligible compared to $n$. Their performance under ``many instruments" scenarios, where $K$ is allowed to grow proportionally with $n$, remains unclear.

The literature on inference with  ``many weak IVs" has grown substantially in recent years.  
Within the many-instrument framework introduced by \citet{bekker1994alternative}, \citet{anatolyev2011specification} proposed a modified Anderson–Rubin (AR) test that allows the instrument-to-sample-size ratio $K/n \rightarrow \mu$ for some constant $0 <\mu   < 1$, under the assumption of conditional homoskedasticity. However, homoskedasticity rarely holds in practice, and inference procedures that accommodate heteroskedasticity are of greater practical relevance. To address this, several heteroskedasticity-robust testing methods have been developed.  For instance, \citet{chao2014testing} introduced a jackknife-based AR test and considered a heteroskedasticity-robust version of the Fuller estimator \citep{fuller1977some}, which was also discussed in \citet{hausman2012instrumental}.   Their procedure assumes that  $K/n$ remains bounded and imposes restrictions on the growth rate of  $K$  relative to the square of the concentration parameter. Similarly, \citet{crudu2021inference} proposed a test under comparable assumptions, while \citet{mikusheva2022inference} introduced a related statistic with a refined variance estimator. Although the latter improves performance in certain scenarios, both approaches still restrict the growth of $K$. A major limitation of these procedures is the requirement that  $K <n$. Their finite-sample performance tends to deteriorate as $K$  approaches $n$.

Recognizing this drawback, recent literature has extended the analysis to ``very many IVs"  scenarios, allowing the number of IVs to be greater than the number of observations.
For instance, \citet{belloni2012sparse} proposed the Sup Score test (henceforth BCCH), specifically  designed to handle ``very many IVs" scenarios. Although the BCCH test uniformly controls size across a range of designs,  this robustness comes at the expense of substantial power loss, a limitation explicitly acknowledged by \cite{belloni2012sparse}. The poor power performance of the BCCH test likely arises
from the sparsity assumption in the first-stage model and the use of a conservative critical value that grows with  $K$, thereby reducing the rejection probability under the alternative hypotheses.
Another approach involves regularization via ridge methods.  \citet{carrasco2016regularization} (henceforth CT) introduced a ridge-regularized AR statistic  suitable for high-dimensional IV models. However, this procedure lacks robustness to arbitrary heteroscedasticity in the error terms. To address this, \citet{dovi2024ridge} proposed a test (henceforth RJAR), which integrates jackknife and ridge regularization to construct a heteroskedasticity-robust inference procedure under arbitrarily weak identification. 
However, ridge-based procedures often require the computation of high-dimensional projection matrices and the solution of optimization problems to determine suitable regularization parameters. Moreover, their theoretical properties under alternative hypotheses remain relatively unexplored.

In this paper, we propose three inference procedures in high-dimensional setting with weak instruments. Our main contributions are summarized below:
\begin{itemize}
	\item Our procedures accommodate scenarios where the number of IVs exceeds the sample size. Moreover, they are robust to identification strength of IVs, and remain valid in the presence of heteroskedastic errors.
	\item We propose a new quadratic-type test, which is powerful under dense IV configurations and, in contrast to the RJAR method \citep{dovi2024ridge}, avoids the need for selecting a ridge-regularization parameter, thereby offering improved computational efficiency. We refine the BCCH procedure \citep{belloni2012sparse} by explicitly deriving the limiting null distribution of the BCCH test statistic. This refinement enables the use of less conservative critical values, effectively addressing the size and power distortions observed in the BCCH test as $K \to \infty$, and thereby improving overall inferential accuracy.    
	\item We establish the asymptotic independence between the quadratic-type and maximum-type statistics. This theoretical result supports aggregating their respective $p$-values using Fisher's  method, resulting in a combination test  with enhanced power that remains effective without prior knowledge of the underlying sparsity.
	
	
\end{itemize}

The remainder of the paper is organized as follows. In Section~\ref{section2}, we introduce the proposed inference procedures tailored to both dense and sparse IV settings, and develop a Fisher-combination test to handle scenarios with unknown IV sparsity structures. Section~\ref{section3} establishes the corresponding asymptotic theory. Section~\ref{section4} reports results from simulation studies that evaluate the finite-sample performance of the proposed methods, while Section~\ref{section5} presents an empirical application using a real data set. Conclusions and discussions are presented in Section~\ref{section6}. Additional simulation results and all proofs are provided in the Supplementary Material.

\textbf{Notation.}  
For a vector $\boldsymbol{a}=\left(a_{1},\ldots, a_{m}\right)^{\top}\in \mathbb{R}^{m}$, we denote its $\ell_{q}$ norm as $\|\boldsymbol{a}\|_{q}=\left(\sum_{\ell=1}^{m}|a_{\ell}|^{q}\right)^{1/q}, \ 1\leq q< \infty$ and $\|\boldsymbol{a}\|_{\infty}=\max_{1\leq \ell \leq m}|a_{\ell}|$.
For any integer $m$, we define $[m] = \{1, \ldots, m\}$. The sub-Gaussian norm  of a scalar random variable $X$ is defined as $\|X\|_{\psi_{2}}=\inf{\{c>0: \mathbb{E}\exp(X^{2}/c^{2})\leq 2\}}$.
        The sub-Exponential norm of a random variable $X$ is defined as $\|X\|_{\psi_1}=\inf\{t>0: \mathbb{E}\exp(|X|/t) \leq 2\}$.
For a random vector $\boldsymbol{X} \in \mathbb{R}^m$, we define its sub-Gaussian norm as 
$\|\boldsymbol{X}\|_{\psi_2} = \sup_{\|\boldsymbol{c}\|_2 = 1} \|\boldsymbol{c}^\top \boldsymbol{X} \|_{\psi_2}$.
Furthermore, we use $\bf{I}_{K}$,  $\boldsymbol{1}_{K}$ and $\boldsymbol{0}_{K}$ to denote the identity matrix in $\mathbb{R}^{K \times K}$, a vector of dimension $K$ with all elements being $1$ and all elements being $0$, respectively.
For a matrix ${\bf{A}} = [A_{jk}]$, ${\bm A}_j$ represents the $j$-th row of ${\bf{A}}$. 
In addition, we use $\lambda_{\min}({\bf{A}})$ and $\lambda_{\max}({\bf{A}})$ to denote the minimal and maximal eigenvalues of ${\bf{A}}$, respectively. 
We define $\mathrm{tr}({\bf{A}})$ as the trace of matrix ${\bf{A}}$.
We use
\(\mathrm{rank}({\bf{Z}})\) to denote the rank of the matrix \({\bf{Z}}\), and  $\mathrm{corr}(X_1,X_2)$ to  denote the correlation coefficient between  variables $X_1$ and $X_2$.
We use $|\mathcal{A}|$ to denote the cardinality of a set $\mathcal{A}$.
 Let $\xrightarrow{d}$ denote convergence in distribution. 
For two positive sequences $\{a_n\}_{n \geq 1}$, $\{b_n\}_{n \geq 1}$, we write $a_n = O(b_n)$ if there exists a positive constant $C$ such that $a_n \leq C \cdot b_n$, and we write $a_n = o(b_n)$ if $a_n / b_n \to 0$. Furthermore, if $a_n = O(b_n)$ is satisfied, we write $a_n \lesssim b_n$. If $a_n \lesssim b_n$ and $b_n \lesssim a_n$, we will write it as $a_n \asymp b_n$ for short.
In addition, let $a_n = O_{{P}}(b_n)$ denote $\Pr(|a_n / b_n| \leq c) \to 1$ for some constant $c < \infty$. Let $a_n = o_{{P}}(b_n)$ denote $\Pr(|a_n / b_n| > c) \to 0$ for any constant $c > 0$.

\section{Inference procedures}\label{section2}

In this paper, we consider the following  linear IV regression model: 
\begin{equation}\label{ivmodel}
\begin{aligned}
	Y &= X\beta + \varepsilon, \\
	X &= \boldsymbol{Z}^\top\boldsymbol{\Pi} + v, 
\end{aligned}
\end{equation}
where \( Y \in \mathbb{R} \) is the response variable, \( X \in \mathbb{R} \) is a scalar endogenous variable, and  \( \boldsymbol{Z} \) is a \( K\)-dimensional vector of IVs. Without loss of generality, we assume that $X$ and $\boldsymbol{Z}$ are centered. Our analysis  focuses on settings where $K$ is diverging. The parameter vector \( \boldsymbol{\Pi} \in \mathbb{R}^K\) captures the relationship between $X$ and $\bm{Z}$, while the scalar parameter \(\beta\) characterizes the structural relationship between $X$ and $Y$.
Denote \( \varepsilon \) and \( v \) as  the structural  and the first-stage error terms, respectively. 
 We assume that the instrument vector  $\bm{Z}$ satisfies $\mathbb{E}(\bm{Z} \varepsilon)=\bm{0}_K$ and $\mathbb{E}(\bm{Z} v)=\bm{0}_K$. As with many authors including \cite{belloni2012sparse, crudu2021inference, dovi2024ridge}, we simplify our analysis by excluding exogenous variables from the structural equation. In Section \ref{sectiona} of the Supplementary Material, we extend the setting to explicitly account for the presence of exogenous variables. We are interested in the inference of the coefficient \( \beta \).  To this end, we consider testing the hypothesis:
\begin{equation}\label{test}
	H_0: \beta = \beta_0 \quad \text{versus} \quad H_1: \beta \neq \beta_0, 
\end{equation}
where $\beta_0$ is a known constant.

We assume that the data $\{X_i, \boldsymbol{Z}_i, Y_i\}_{i=1}^n$ are independent and identically distributed (i.i.d.) copies of $\{X, \boldsymbol{Z}, Y\}$. Let $\boldsymbol{X}=(X_1,\ldots,X_n)^{\top} \in \mathbb{R}^n$,  ${\bf Z}=(\boldsymbol{Z}_1,\ldots,\boldsymbol{Z}_n)^{\top} \in \mathbb{R}^{n \times K}$ and $\boldsymbol{Y}=(Y_1, \ldots,Y_n)^{\top} \in \mathbb{R}^n$.   Under the null hypothesis $H_0: \beta=\beta_0$, we define the structural error as
$e_i  :=Y_i-X_i \beta_0$. 
The core idea of weak-identification-robust AR statistic is that the structural error $e_i$ is uncorrelated with the instruments  under the null hypothesis, i.e., $\mathbb{E}(\boldsymbol{Z}_i e_i)=\boldsymbol{0}_K$. 
Under $H_0$,  the AR  statistic asymptotically follows a chi-square distribution with $K$ degrees of freedom. However, conventional AR tests perform poorly when $K \rightarrow \infty$. A main reason for the poor performance is due to the singularity
of ${\bf Z}^\top {\bf Z}$ in high-dimensional settings. To solve this critical issue, \cite{dovi2024ridge} adopted ridge-regularizitian approach. However, in the following, we aim to show that it is actually not necessary to adopt ridge-regularizitian.

Now we first revisit the ridge-regularized jackknifed Anderson-Rubin test (RJAR) \citep{dovi2024ridge}, which is defined as:
\begin{equation}
	\mathrm{RJAR}_{\gamma_n} := \frac{1}{ \sqrt{r_n \widehat{\Phi}_{\gamma_n}}} 
	\sum_{i=1}^{n} \sum_{j \ne i}^n P_{ij}^{\gamma_n} e_i e_j, 
\end{equation}
where $r_n := \mathrm{rank}({\bf Z})$, $ \widehat{\Phi}_{\gamma_n} := 2 r_n^{-1} \sum_{i=1}^{n} \sum_{j \ne i}^n (P_{ij}^{\gamma_n})^2 e_i^2 e_j^2, $
and ${\bf {P}}^{\gamma_n} := {\bf Z}({\bf Z}^\top {\bf Z} + \gamma_n \bf{I}_{K})^{-1} {\bf Z}^\top$ is the ridge-regularized projection matrix with regularization parameter $\gamma_n$. By incorporating ridge regularization, the RJAR statistic accommodates ``very many IVs" scenarios. However,  its implementation involves computing a ridge-regularized projection matrix and selecting the regularization parameter through optimization (see their equation (7)). These steps can be computationally intensive, particularly when  $K \gg n$, thereby limiting the method’s practical applicability. 
To address these limitations, we propose a jackknifed Anderson-Rubin (JAR) test without ridge-regularizitian:
\begin{equation}\label{quadratic}
	\mathrm{JAR}_{n}:= \frac{1}{ \sqrt{r_n \widehat{\Phi}_n}} 
	\sum_{i=1}^{n} \sum_{j \ne i}^n P_{ij} e_i e_j, 
\end{equation}
where $r_n=\mathrm{rank}({\bf Z})$, $ \widehat{\Phi}_n := 2 r_n^{-1} \sum_{i=1}^{n} \sum_{j \ne i}^n (P_{ij})^2 e_i^2 e_j^2, $
and ${\bf {P}} := {\bf Z}{\bf Z}^\top$ is the projection matrix without regularization parameter $\gamma_n$. It is easy to know that the above $\mathrm{JAR}_{n}$ test statistic can also be rewritten as follows:
\begin{equation*}
	\mathrm{JAR}_{n}:= \frac{\sum_{i=1}^{n} \sum_{j \ne i}^n e_i e_j\bm Z_i^\top\bm Z_j}{ \sqrt{2\sum_{i=1}^{n} \sum_{j \ne i}^n e_i^2 e_j^2(\bm Z_i^\top\bm Z_j)^2}}.	 
\end{equation*}

Our proposed statistic $\mathrm{JAR}_{n}$ can be viewed as a simplified variant of the $\mathrm{RJAR}_{\gamma_n}$ statistic by replacing the ridge-regularized projection matrix ${\bf P}^{\gamma_n}$ with the  simpler  matrix ${\bf P}$. We simply set the matrix $({\bf Z}^\top {\bf Z} + \gamma_n \bf{I}_{K})^{-1}$ in ${\bf {P}}^{\gamma_n}$ as an identity matrix $\bf{I}_{K}$. Since now there is no regularization parameter $\gamma_n$ and no need to compute inverse matrix, our test $\mathrm{JAR}_{n}$ is very easy to implement. 
As demonstrated in the Section \ref{section4.2}, our proposed statistic not only significantly reduces computational complexity but also yields improved power relative to existing approaches.
According to Theorem \ref{theorem1}, the test JAR rejects $H_0$ at the significant level $\alpha$ if $\mathrm{JAR}_{n}\geq z(\alpha)$, where $z(\alpha)$ denotes the upper $\alpha$-quantile of  standard Gaussian distribution $N(0, 1)$.

The above $\mathrm{JAR}_{n}$ statistic and also the $\mathrm{RJAR}_{\gamma_n}$ statistic are both quadratic-type test statistics. These test statistics usually perform well when the alternatives are dense \citep{chen2010two, srivastava2013two,10.1214/17-AOS1573,chen2019two}. However, when the vector $\mathbb{E}(\bm Z_i e_i)=\mathbb{E}(\bm Z_i X_i)(\beta-\beta_0)$ is extremely sparse, they may perform not well \citep{xu2016adaptive, yu2024fisher}. For sparse scenarios, maximum-type test statistics should be considered \citep{tony2014two,cao2018two}. 

Now we consider the maximum-type statistic introduced by \cite{belloni2012sparse}:
\begin{equation}\label{maximum}
	M_n = \max_{1 \leq k \leq K} \left|\frac{S_{nk}}{\hat{\sigma}_k}\right|,
\end{equation}
where
$S_{nk} = {n}^{-1/2} \sum_{i=1}^{n} e_i Z_{ik},$
$\hat{\sigma}_k^2 = n^{-1}\sum_{i=1}^n e_i^2 {Z}_{ik}^2$ is an estimator of $\sigma_k^2 = \mathbb{E} (e_i^2 {Z}_{ik}^2)$, with $Z_{ik}$ being the $k$-th component of vector $\bm{Z}_i$. The BCCH test rejects the null hypothesis $H_0$ at the significance level $\alpha$ if $M_n > c_{\mathrm{BCCH}}  z\left({ \alpha / 2K}\right) $, with $c_{\mathrm{BCCH}}=1.1$ and $z\left({ \alpha / 2K}\right)$ being the upper $(\alpha / 2K)$-quantile  of $N(0, 1)$. 
However, the BCCH test tends to be substantially conservative in practice, as noted by \cite{belloni2012sparse, dovi2024ridge, lim2024dimension}. This conservative behavior likely arises because the critical values employed by the BCCH test become excessively stringent when $K \rightarrow \infty$. Actually \cite{belloni2012sparse} adopted the Bonferroni adjustment to determine the critical value. However the Bonferroni adjustment is well known to be conservative \citep{narum2006beyond}.

Motivated by this limitation, we derive the limiting null distribution of the  squared BCCH statistic  in Theorem \ref{theorem3}, allowing for more accurate critical values and thereby improving the test's statistical power. Specifically, under our refined approach, the null hypothesis $H_0$ is rejected at the significant level $\alpha$ if and only if $M_n^2 \ge c(\alpha)$, where $c(\alpha) = 2 (\log K) - \{\log \left(\log K \right)\}+ q_\alpha$, with  $q_\alpha$ being the $(1-\alpha)$-quantile of the Gumbel distribution with the cumulative distribution function $G(x)=\exp\left\{-{\pi}^{-1/2}\exp(-x/2)\right\}$, that is, $q_\alpha = -(\log\pi) - 2 \log \left\{\log (1 - \alpha)^{-1}\right\}$. As demonstrated in Section \ref{section4.3}, $c_{\mathrm{BCCH}}  z\left({ \alpha / 2K}\right)$ is generally larger than $\sqrt{c(\alpha)}$. Thus our refinement improves the power performance of the BCCH procedure.

The quadratic-type statistic $\mathrm{JAR}_{n}$ demonstrates substantial power against dense alternatives, where maximum-type statistics often underperform. 
 In contrast, the maximum-type statistic $M_n^2$ is particularly effective under sparse alternatives, see \citet{10.1214/12-AOS993,cai2013two,feng2022asymptotic,chen2024high} for further discussion.  To harness the complementary strengths of both approaches, we adopt Fisher's method \citep{fisher1970statistical} to construct a combined test. The proposed Fisher-combination test leverages the asymptotic independence between $\mathrm{JAR}_{n}$ and $M_n^2$,  ensuring sensitivity across a wide range of alternative hypotheses. Specifically, the combined test statistic is defined as:
\begin{equation}\label{fisher}
	F_{n} = -2\left( \log p_J\right) - 2 \left(\log p_M\right), 
\end{equation}
where $p_J = 1 - \Phi \left(\mathrm{JAR}_{n}\right) $ and $p_M = 1 - G \left[ M_n^2 - 2 (\log K )+ \left\{ \log \left(\log K\right) \right\} \right]$
are the $p$-values associated with the test statistics $\mathrm{JAR}_{n}$ and $M_{n}^2$, respectively. Here $\Phi(\cdot)$ and $G(\cdot)$ denote the cumulative distribution functions of the standard normal and Gumbel distributions, respectively. 
Denote $l\left({\alpha}\right)$ as the upper $\alpha$-quantile of the  chi-square
distribution with four degrees of freedom,  i.e.,  $\chi^2_4$. The null hypothesis is rejected at the significance level $\alpha$ if
$F_{n} \geq l\left(\alpha\right).$
A key advantage of the proposed combination procedure lies in its robustness across both sparse and dense alternatives, without the need for tuning parameters, 
thereby facilitating straightforward implementation. Moreover, Fisher's method is known to achieve asymptotic optimality in terms of Bahadur efficiency, as established in \citet{littell1971asymptotic, littell1973asymptotic}.

\section{Theoretical results} \label{section3}
This section presents the main theoretical results for the three proposed tests: the quadratic-type, maximum-type, and Fisher-combination tests.  Section \ref{subsection3.1} introduces the required notation and assumptions. Section \ref{subsection3.2} derives the asymptotic properties of the quadratic-type statistic $\mathrm{JAR}_{n}$. Section \ref{subsection3.3} establishes the theoretical results for the maximum-type statistic $M_n^2$. Section \ref{subsection3.4}  establishes the asymptotic independence of $\mathrm{JAR}_{n}$ and $M_n^2$ under the null hypothesis and analyzes the power properties of the Fisher-combination test.

\subsection{Assumptions} \label{subsection3.1}
\begin{assumption}\label{assumption1}
	It holds that
	\begin{itemize}
		\item[(i)]  There exists a positive constant $c_0$ such that $\| \bm{Z} \|_{\psi_2} \leq c_0$,  $\| \varepsilon \|_{\psi_2} \leq c_0$, and $\| v \|_{\psi_2} \leq c_0$;
		
		
		\item[(ii)]   The random error $\varepsilon$ satisfies $ \mathbb{E}(\varepsilon^2 | \bm{Z}) < \infty$.
		
	\end{itemize}
\end{assumption}

\begin{assumption}\label{assumption2}
    Assume that $\operatorname{tr}(\boldsymbol{\Sigma}_{\bm Z}^4) = o\{\operatorname{tr}^2(\boldsymbol{\Sigma}_{\bm Z}^2)\}$ as $(n, K) \to \infty$, where $\bm{\Sigma}_{\bm Z} = \mathbb{E}\left(\bm{Z}\bm{Z}^\top\right)$.
\end{assumption}

Assumption \ref{assumption1}(i) is a standard sub-Gaussian condition on the error term commonly imposed in high-dimensional linear models \citep{papaspiliopoulos2020high}, and similar assumptions on instrumental variables appear in \cite{zhu2018sparse, gold2020inference, fan2024heteroscedasticity}.
Assumption \ref{assumption1}(ii) imposes mild regularity on the structural error, accommodating conditional heteroscedasticity.
Assumption \ref{assumption2} is a mild condition commonly found in the literature \citep{chen2010two,zhong2011tests,guo2016tests,10.1214/17-AOS1573}. 
%

Before proceeding to the next assumption, we introduce several useful notation used throughout our analysis. 
Denote the standardized version of the variable $S_{nk}$ as  $\bar{S}_{nk} = \{\mathrm{Var}(S_{nk})\}^{-1/2} \left\{ S_{nk} - \mathbb{E}(S_{nk})\right\}$. Now define  $\bm{\bar{S}} = ( \bar{S}_{n1}, \dots, \bar{S}_{nK} )^\top \in \mathbb{R}^{K}$, and $\mathbf{\Lambda}^* = \mathbb{E} (\bm{\bar{S}} \bm{\bar{S}}^\top)=(\Lambda_{jk}^*)_{1\leq j,k \leq K}$.
For a given $0 < r < 1$, define the set 
$ \mathcal{HC}_k( r) = \{1 \leq j \leq K : |\Lambda_{kj}^*| \geq r\} $  
which collects the set of indices $j$ for which the components $\bar{S}_{nj}$ are highly correlated with $\bar{S}_{nk}$,   given a $k \in [K]$.
Additionally, define  
$ \mathcal{W}( r) = \{1 \leq k \leq K : \mathcal{HC}_k( r) \neq \emptyset\} $  
which contains all indices $k$ for which the component $\bar{S}_{nk}$ is highly correlated with at least one other component in $\bm{\bar{S}}$.
Furthermore, for any $c > 0$, let $ s_k( c) = |\mathcal{HC}_k\{ (\log K)^{-1-c}\}|$  be the number of variables $\bar{S}_{nj}$ highly correlated with $\bar{S}_{nk}$ at the threshold $ (\log K)^{-1-c}$, given a $k \in [K]$.

\begin{assumption} \label{assumption4}
	\textit{$( \log K )^{13} = o(n)$}.
\end{assumption}



\begin{assumption} \label{assumption8}
	\textit{We make the following assumptions:
		\begin{itemize}
			\item[(i)] There exists a constant $0 < r < 1$, such that $|\mathcal{W}( r)| = o(K)$;
			\item[(ii)]There exists a subset $\bm{\Upsilon} \subset \{1,  \ldots, K\}$ satisfied $|\bm{\Upsilon}| = o(K)$. For some constant $c_2 > 0$ and all $\gamma > 0$, such that $\max_{k \in [K], k \notin {\bm{\Upsilon}}} s_k( c_2) = o(K^\gamma) $.
	\end{itemize}}
\end{assumption}

Assumptions \ref{assumption4} is mild and commonly adopted in high-dimensional settings \citep{ li2024power}. In particular, it also  allows the number of instruments $K$ to grow exponentially with the sample size. 
Assumption \ref{assumption8} (i) imposes mild conditions on the dependence structure among the components  of $\bm{\bar{S}}$, and similar conditions have been considered in \cite{li2024power}.
Assumption \ref{assumption8} (ii) provides a more general requirement than the commonly imposed boundedness condition on the eigenvalues of the correlation matrix. 
 Following \cite{tony2014two} and \cite{cao2018two}, the limiting distribution of the maximum-type statistic can be established under alternative conditions:  $\max_{1 \leq  j \neq k \leq K}|\Lambda^*_{jk}| \leq r_1$  and $\max_{j \in [K]}\sum_{k=1}^K\Lambda^{*2}_{jk} \leq r_2,$ for
some constants $0<r_1<1$ and $r_2>0$. Notably, Assumption \ref{assumption8} (ii) encompasses these as special cases when  $s_j(1)=C(\log K)^2$, as discussed in \cite{li2024power}.

\subsection{Theoretical results for the quadratic-type test } \label{subsection3.2}
In this subsection, we establish the limiting null distribution and analyze the power property of the proposed test statistics $\mathrm{JAR}_{n}$.

\begin{theorem}\label{theorem1}
	\textit{Suppose that Assumption \ref{assumption1}-\ref{assumption2} hold, then under the null hypothesis $H_0$, we have}
	\begin{equation}
		\mathrm{JAR}_{n} \xrightarrow{d} N(0,1) \quad \text{as } (n, K) \to \infty.
	\end{equation}
\end{theorem}

Define $T_n = \{n(n-1)\}^{-1} \sum_{i=1}^n\sum_{j\neq i}^{n} e_i e_j \boldsymbol{Z}_{i}^\top \boldsymbol{Z}_{j}$, and let  $\widehat\Omega=2
(n-1)^{-2} \sum_{i=1}^n\sum_{j\neq i}^{n} e_i^2 e_j^2 (\boldsymbol{Z}_{i}^\top \boldsymbol{Z}_{j})^2$ denote an estimator of $\Omega=2 \mathbb{E}\{e_i^2 e_j^2 (\boldsymbol{Z}_i^\top \boldsymbol{Z}_j)^2\}$. We then have $\mathrm{JAR}_{n}=nT_n/\sqrt{\widehat\Omega}$. 
Next, we investigate the limiting distribution of $\mathrm{JAR}_{n}$ under a class of local alternatives. Denote the signal strength vector as  $\bm{\zeta}=\boldsymbol{\Pi}(\beta-\beta_0)$, and the following family of local alternatives are defined as:
\begin{equation}
	\mathscr{L}_{1}(\bm{\zeta}) = \left\{ \bm{\zeta} \in \mathbb{R}^{K} \,\middle|\, 
	\begin{array}{l}
		\bm{\zeta}^\top \boldsymbol{\Sigma_Z} \bm{\zeta} = o(1) \\
		\bm{\zeta}^\top \boldsymbol{\Sigma_Z}^3 \bm{\zeta} = o\left\{ \frac{\operatorname{tr}(\boldsymbol{\Sigma_Z}^2)}{n} \right\}
	\end{array} 
	\right\}.
\end{equation}
Similar definitions of local alternative sets have been utilized previously  in the literature; see, for example, \cite{zhong2011tests, 10.1214/17-AOS1573, guo2022conditional}. The class $\mathscr{L}_1(\bm{\zeta})$ characterizes alternatives under which the deviation of $\bm{\zeta}$ from $\bm{0}_K$ is small. We have the following theorem.

\begin{theorem}\label{theorem2}
	\textit{Under the conditions of Theorem \ref{theorem1}, for $\bm{\zeta} \in \mathscr{L}_{1}(\bm{\zeta})$, we have}
	\begin{equation}
		\mathrm{JAR}_{n} - \frac{  n \bm{\zeta}^\top \boldsymbol{\Sigma_Z}^2 \bm{\zeta}}{\sqrt{\widehat{\Omega}}} \xrightarrow{d} N(0,1) \quad \text{as } (n, K) \to \infty.
	\end{equation}
\end{theorem}

Theorem \ref{theorem2} indicates that the asymptotic power of the test statistic $\mathrm{JAR}_{n}$ under the local alternatives $\mathscr{L}_1(\bm{\zeta})$ is given by
\begin{equation} \label{quadratic.power}
	\phi_{1n} = \Phi\left(-z_{\alpha} + \dfrac{n \bm{\zeta}^\top \boldsymbol{\Sigma_Z}^2 \bm{\zeta}}{\sqrt{\widehat{\Omega}}}\right).
\end{equation}
Equation (\ref{quadratic.power}) implies that the proposed test has nontrivial power as long as the signal-to-noise ratio $n \bm{\zeta}^\top \boldsymbol{\Sigma_Z}^2 \bm{\zeta} / \sqrt{\widehat{\Omega}}$ does not vanish to $0$ as $(n, K) \rightarrow \infty$.

Finally, Proposition \ref{proposition1}  shows that 
$\widehat{\Omega}$ is a ratio consistent estimator of  $\Omega$.
\begin{proposition}\label{proposition1}
	\textit{Under the conditions  in Theorem \ref{theorem1}, we have
		$$
		\dfrac{\widehat{\Omega}}{\Omega}\rightarrow 1,
		$$
		where $\Omega=2 \mathbb{E}\{e_i^2 e_j^2 (\boldsymbol{Z}_i^\top \boldsymbol{Z}_j)^2\}, \widehat{\Omega}={2}(n-1)^{-2}\sum_{i\neq j}e_i^2 e_j^2 (\boldsymbol{Z}_i^\top \boldsymbol{Z}_j)^2$.}
\end{proposition}

\subsection{Theoretical results for the maximum-type test}\label{subsection3.3}
In this subsection, we derive the asymptotic null distribution and provide the asymptotic power property for the maximum-type  statistic $M_n^2$ defined in (\ref{maximum}).

\begin{theorem}\label{theorem3}
	Suppose that Assumptions \ref{assumption1}, \ref{assumption4}, \ref{assumption8} hold,
	then under the null hypothesis $H_0$, for a given $t \in \mathbb{R}$,  we have
	\[
	\lim_{(n, K) \to \infty} \sup_{t\in \mathbb{R}}\left| \Pr\left[M_{n}^2 - 2( \log K) + \{\log (\log K)\} \leq t\right] - \exp\left\{-\frac{1}{\sqrt{\pi}} \exp\left(-\dfrac{t}{2}\right)\right\} \right| = 0.
	\]
\end{theorem}
Theorem \ref{theorem3} establishes the asymptotic distribution of $M_n^2$, which enables us to determine the critical value of $M_n^2$, as discussed in Section \ref{section2}. 
We next consider the asymptotic power analysis of the $M_{n}^2$. Define the following parameter space for ${H}_1$,
\begin{equation}
	\mathscr{L}_{2}(c_0) = \left\{ \boldsymbol{\zeta} \in \mathbb{R}^K : \max_{k \in [K]} |\mu_k| \geq \sqrt{c_0 \log K} \right\}, \label{eq3.2}
\end{equation}
where $\mu_k = \sqrt{n} \sum_{j=1}^{K} {\zeta}_j \mathbb{E} (Z_{ij} Z_{ik}) / \sqrt{\sigma_k^2}, $
 ${\zeta_j}={\Pi_j}(\beta-\beta_0)$ is the $j$-th component of the vector $\boldsymbol{\zeta}=\boldsymbol{\Pi}(\beta-\beta_0)$, and  $c_0$ is a positive constant.

\begin{theorem}\label{theorem4}
	\textit{Suppose that the conditions in  Theorem \ref{theorem3} hold, then for some $\epsilon_0 > 0$,}
		\begin{equation}
			\lim_{(n, K) \to \infty} \inf_{\bm{\zeta} \in \mathscr{L}_{2} (2+\epsilon_0)} \Pr \{M_{n}^2 > c(\alpha)\} = 1,
		\end{equation}
		where $c(\alpha) = 2 (\log K) - \{\log (\log K)\} + q_{\alpha}$ and   $q_{\alpha} = -(\log \pi) - 2[ \log\{ \log (1 - \alpha)^{-1}\}]$.
\end{theorem}
Theorem \ref{theorem4} demonstrates that our maximum-type test is powerful provided that the separation rate is of order $\sqrt{\left\{(2+\epsilon_{0})(\log K) \right\} / n}$ for some $\epsilon_0 > 0$ when $K \rightarrow \infty$. This rate of signal separation is well established in existing studies,  such as  \cite{tony2014two, zhang2017simultaneous, ma2021global}.

\subsection{Theoretical results for the Fisher-combination test}\label{subsection3.4}
In this subsection, we present the joint asymptotic distribution of the statistics $\mathrm{JAR}_{n}$ and $M_n^2$ under the null hypothesis. Furthermore, we establish theoretical properties of the Fisher-combination statistic $F_n$ defined in (\ref{fisher}).

\begin{theorem}\label{theorem5}
	Under Assumptions \ref{assumption1}-\ref{assumption8}, and the null hypothesis $H_0$, we have
	\begin{equation}
		\Pr[\mathrm{JAR}_{n} < x, M_n^2 - 2 (\log K) + \{\log \left(\log K \right)\} < y] \xrightarrow{d} \
		{\Phi}(x)\cdot G(y),
	\end{equation}
	for any \( x, y \in \mathbb{R} \), as \( n, K \to \infty \). Here, 
	\( \Phi(\cdot) \) and $	G(\cdot)	$
	  denote the cumulative distribution functions of the standard normal and  Gumbel distributions, respectively.
\end{theorem}
Theorem \ref{theorem5} implies the asymptotic independence between the statistics $\mathrm{JAR}_{n}$ and $M_n^2$. This asymptotic independence justifies the validity of combining these statistics using Fisher's method.  Corollary \ref{theorem6} shows that the test procedure $F_n$ can control the size asymptotically, and Corollary \ref{theorem7} provides its asymptotic power property.

\begin{corollary}\label{theorem6}
	Under the conditions in Theorem \ref{theorem5} and the null hypothesis $H_0$, the Fisher-combination statistic $F_n$ satisfies
	\begin{equation}
		\Pr{\left\{ F_n \geq l\left(\alpha\right) \right\}} \to \alpha \quad \text{as } n, K \to \infty,
	\end{equation}
where $l\left(\alpha\right)$ is the upper $\alpha$-quantile of $\chi_4^2$ distribution.
\end{corollary}

\begin{corollary}\label{theorem7}
	Suppose that  assumptions of Theorem \ref{theorem2} and Theorem  \ref{theorem4} hold. Then, the Fisher-combination test $F_n$ achieves consistent asymptotic power under the alternative hypothesis $H_1$:
	\begin{equation}
		\inf_{\bm{\zeta} \in \mathscr{L}_{1}\cup\mathscr{L}_{2}(c)} \Pr{\left\{F_{n} \geq l\left(\alpha\right)\right\}}\to 1 
		\quad \text{as } n, K \to \infty.
	\end{equation}
\end{corollary}
Corollary \ref{theorem7} shows that the Fisher-combination test is consistently powerful without requiring prior knowledge of the underlying sparsity pattern.

\section{Simulation studies} \label{section4}
In this section, we conduct  simulation studies to evaluate the finite-sample performance of the proposed methods. 
For a comprehensive comparison, we include two existing methods, the RJAR test proposed by \cite{dovi2024ridge} and the BCCH test introduced by \cite{belloni2012sparse}, alongside our three proposed tests: the quadratic-type test defined in (\ref{quadratic}) (henceforth $\mathrm{JAR}$), the maximum-type test $M_n^2$ defined in (\ref{maximum}) (henceforth BCCH$\_$Asy), and the Fisher-combination test defined in (\ref{fisher}) (henceforth Fisher). 
Section~\ref{section4.1} reports the size and power performance of the five procedures.
Section~\ref{section4.2}  compares the computation time of our proposed $\mathrm{JAR}$ and RJAR test.
Section~\ref{section4.3} presents a comparison between the BCCH-based critical value and our refined threshold for the maximum-type statistic in (\ref{maximum}).
We consider the following model:
\begin{equation}
\begin{gathered}
	Y_i=X_i \beta+\varepsilon_i, \\
	X_i=\boldsymbol{Z}_i^{\top} \boldsymbol{\Pi}+v_i,
\end{gathered}
\end{equation}
for  $i=1,\ldots,n$. Here,  $Y_i$ is the response variable, \(X_i \in \mathbb{R}\) is the scalar endogenous regressor,  \(\boldsymbol{Z}_i \in \mathbb{R}^{K}\) denotes the vector of instrumental variables.  \(\beta\) is the parameter of interest. 
We examine a correlated Gaussian structure of $\bm{Z}_i$, generating independent and identically Gaussian vectors with zero mean  and correlation structure defined as  $\mathrm{corr} \left(Z_{i l}, Z_{i m}\right)=0.6^{|l-m|}$.
The error terms $\varepsilon_i$ and $v_i$ are given by 
\begin{equation}
	\begin{aligned}
		\varepsilon_i=&\left(\sigma_{\varepsilon}+a_0 Z_{i1} \cdot e_i^a\right)\eta_{1i}, \\
		v_i=&\sigma_v \eta_{2i},
	\end{aligned}
\end{equation}
where $\sigma_{\varepsilon}^2=2, \sigma_v^2=1$, and  
\(\eta_{1i}\) and \(\eta_{2i}\) are drawn from \(N(0,1)\) with 
\(\mathrm{Cov}(\eta_{1i}, \eta_{2i}) = 0.6.\) The parameter $a_0$ controls the degree of heteroskedasticity, where $a_0=0$ indicates homoscedasticity, and $a_0\neq0$ corresponds to heteroskedastic error structures. Additionally, $\{ e^a_i \}_{i=1}^{n}$ are independent standard normal variables, and $Z_{i1}$ denotes the first element of $\boldsymbol{Z}_i$.

Furthermore, we examine both sparse and dense scenarios in the first-stage relationship by writing the coefficient vector as  \(\boldsymbol{\Pi} = \tau\,\boldsymbol{\psi}\), where \(\boldsymbol{\psi}\) is a vector consisting of ones and zeros characterizing sparsity or denseness, and the scalar \(\tau\) is chosen to achieve a given concentration parameter \(\mu^2\). Specifically,
\[
\mu^2
= \frac{n\,\boldsymbol{\Pi}^{\top}\,\mathbb{E}(\,\boldsymbol{Z}_i \boldsymbol{Z}_i^{\top}\,)\,\boldsymbol{\Pi}}{\sigma_v^2}
\quad\Longrightarrow\quad
\tau
= \sqrt{\frac{\sigma_v^2\,\mu^2}{n\,\boldsymbol{\psi}^{\top}\,\mathbb{E}(\,\boldsymbol{Z}_i \boldsymbol{Z}_i^{\top}\,)\,\boldsymbol{\psi}}}.
\]
Define $\bm{\psi} = \left(\bm{1}_q^\top, \bm{0}_{K-q}^\top\right)^\top$.
For the sparse scenario, we set $q=0.03K$; 
For the dense scenario, we set $q=0.6K$. 
The nominal significance level is 0.05, and the null hypothesis is evaluated at $\beta_0=1$. 

\subsection{The size and power performance}\label{section4.1}

In this subsection, we utilize the following four examples to conduct a comprehensive comparison of the finite sample performance for the five procedures. Simulation results are based on 300 replications.

\textbf{Example 1.}
\[
\left\{
\begin{array}{l}
	\begin{minipage}[t]{0.9\textwidth}
		{\bf Example 1.1.} We fix $(n,K)=(200,100)$, $a_0=0$ (homoscedasticity), sparse/dense. 
		$\mu^2 \in \{30,180\}$. Under $H_0$: $\beta_0=1$; under $H_1$: $\beta \in \{-1,0,2,3\}$.
	\end{minipage} \\[1ex]
	
	\begin{minipage}[t]{0.9\textwidth}
		{\bf Example 1.2.} We fix $(n,K)=(200,100)$, $a_0=0.5$ (heteroscedasticity), sparse/dense. 
		$\mu^2 \in \{30,180\}$. Under $H_0$: $\beta_0=1$; under $H_1$: $\beta \in \{-1,0,2,3\}$.
	\end{minipage}
\end{array}
\right.
\]

\vspace{1em}

\textbf{Example 2.}
\[
\left\{
\begin{array}{l}
	\begin{minipage}[t]{0.9\textwidth}
		{\bf Example 2.1.} We fix $(n,K)=(200,200)$, $a_0=0$ (homoscedasticity), sparse/dense. 
		$\mu^2 \in \{30,180\}$. Under $H_0$: $\beta_0=1$; under $H_1$: $\beta \in \{-1,0,2,3\}$.
	\end{minipage} \\[1ex]
	
	\begin{minipage}[t]{0.9\textwidth}
		{\bf Example 2.2.} We fix $(n,K)=(200,200)$, $a_0=0.5$ (heteroscedasticity), sparse/dense. 
		$\mu^2 \in \{30,180\}$. Under $H_0$: $\beta_0=1$; under $H_1$: $\beta \in \{-1,0,2,3\}$.
	\end{minipage}
\end{array}
\right.
\]

\vspace{1em}

\textbf{Example 3.}
\[
\left\{
\begin{array}{l}
	\begin{minipage}[t]{0.9\textwidth}
		{\bf Example 3.1.} We fix $(n,K)=(200,300)$, $a_0=0$ (homoscedasticity), sparse/dense. 
		$\mu^2 \in \{30,180\}$. Under $H_0$: $\beta_0=1$; under $H_1$: $\beta \in \{-1,0,2,3\}$.
	\end{minipage} \\[1ex]
	
	\begin{minipage}[t]{0.9\textwidth}
		{\bf Example 3.2.} We fix $(n,K)=(200,300)$, $a_0=0.5$ (heteroscedasticity), sparse/dense. 
		$\mu^2 \in \{30,180\}$. Under $H_0$: $\beta_0=1$; under $H_1$: $\beta \in \{-1,0,2,3\}$.
	\end{minipage}
\end{array}
\right.
\]

\vspace{1em}

\textbf{Example 4.}
\[
\left\{
\begin{array}{l}
	\begin{minipage}[t]{0.9\textwidth}
		{\bf Example 4.1.} We fix $n=200$, $K \in \{100,200,300\}$, $a_0=0$ (homoscedasticity), 
		sparse setting with sparse level $q \in \{1,3,5,7,9\}$, $\mu^2 = 30$. 
		Under $H_1$: $\beta \in \{-1,3\}$.
	\end{minipage} \\[1ex]
	
	\begin{minipage}[t]{0.9\textwidth}
		{\bf Example 4.2.} We fix $n=200$, $K \in \{100,200,300\}$, $a_0=0$ (homoscedasticity), 
		dense setting with dense proportion $\iota \in \{0.2,0.4,0.6,0.8,1.0\}$, 
		and dense level $q=\iota K$, $\mu^2 = 30$. Under $H_1$: $\beta \in \{-1,3\}$.
	\end{minipage}
\end{array}
\right.
\]
\vspace{1mm}

\begin{figure}[h]
	\centering
	\includegraphics[width=1\textwidth]{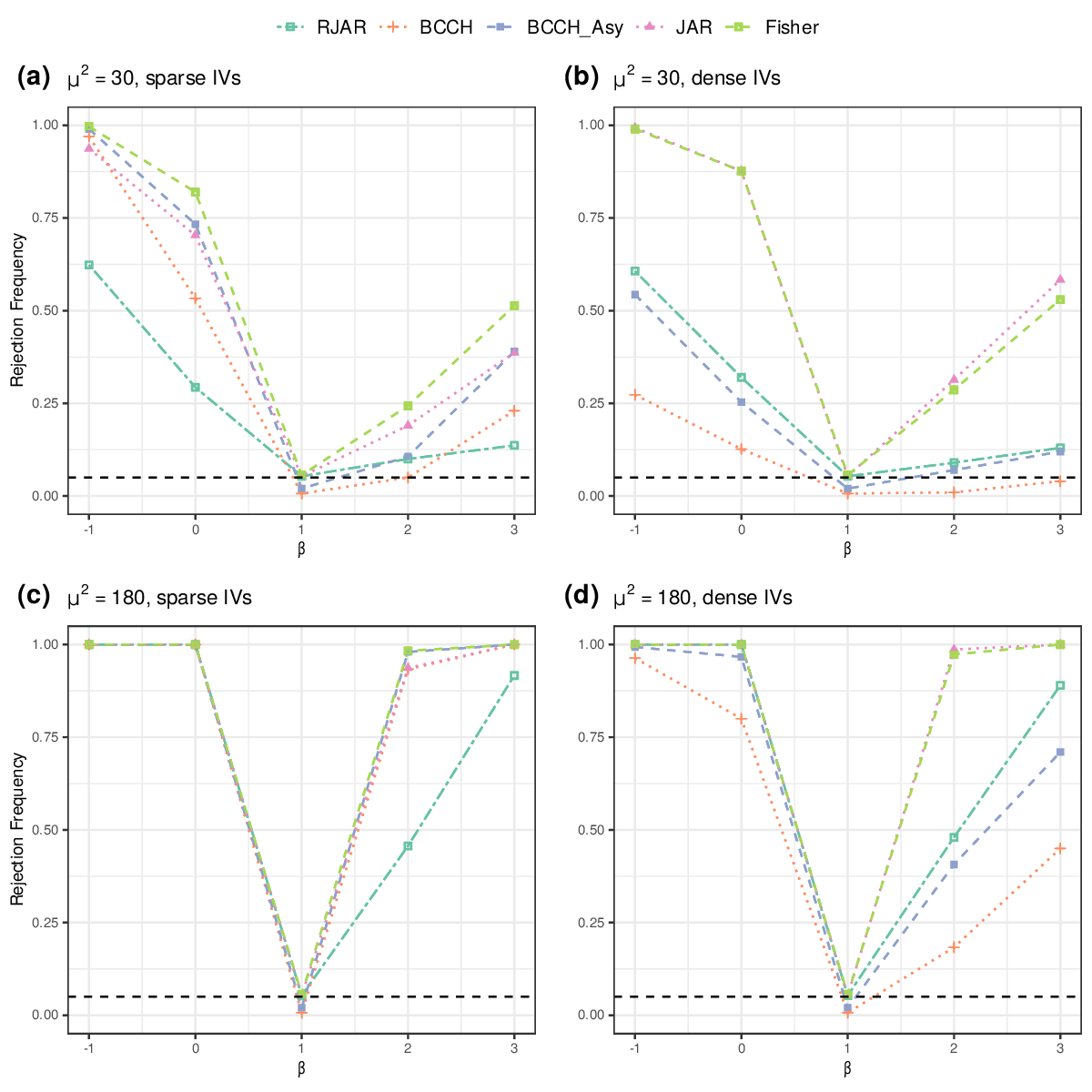}
	\caption{\setstretch{1.5}The rejection frequency curves for Example 1.1 with $K=100$: the RJAR (dark green open square, twodash line), the BCCH (orange plus, dotted line), the BCCH$\_$Asy (purple filled square, dashed line), the $\mathrm{JAR}$ (pink filled triangle, dotted line), and the Fisher (light green square, dashed line). (a)   Sparse scenario with $\mu^2=30$; (b)  dense scenario with  $\mu^2=30$; (c)   sparse scenario with $\mu^2=180$; (d)   dense scenario with $\mu^2=180$.  The nominal test size of 0.05 indicated by the black horizontal line.}\label{figure1}
\end{figure}

\begin{figure}[h]
	\centering
	\includegraphics[width=1\textwidth]{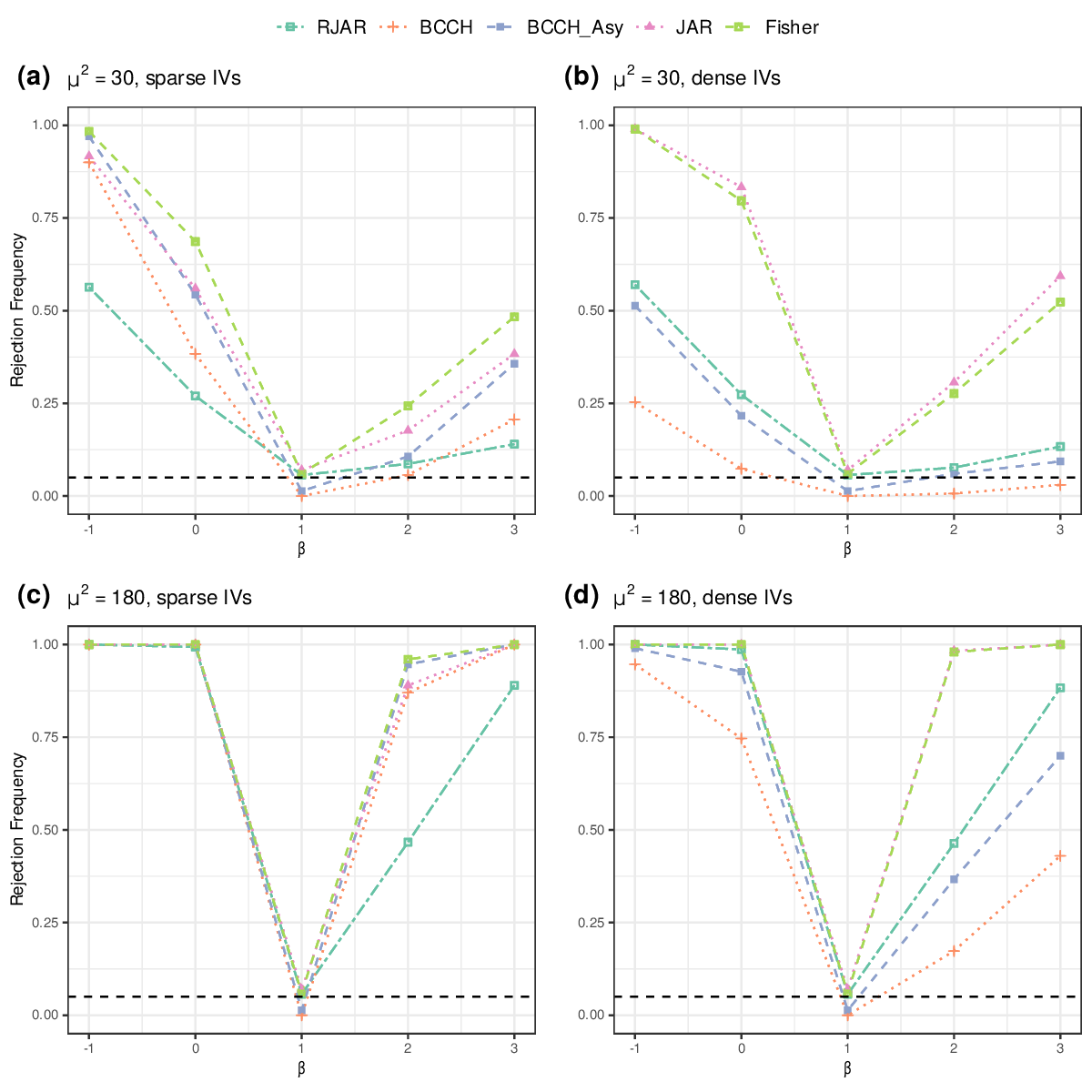}
	\caption{\setstretch{1.5}The rejection frequency curves for Example 1.2 with $K=100$: the RJAR (dark green open square, twodash line), the BCCH (orange plus, dotted line), the BCCH$\_$Asy (purple filled square, dashed line), the $\mathrm{JAR}$ (pink filled triangle, dotted line), and the Fisher (light green square, dashed line). (a)   Sparse scenario with $\mu^2=30$; (b)  dense scenario with  $\mu^2=30$; (c)   sparse scenario with $\mu^2=180$; (d)   dense scenario with $\mu^2=180$.  The nominal test size of 0.05 indicated by the black horizontal line.}\label{figure2}
\end{figure}

To investigate whether the five methods are sensitive to variations in the error-term structure under both sparse and dense scenarios,  as well as to  weak ($\mu^2=30$) and strong ($\mu^2=180$) first-stage identification,  we apply them to Examples 1-3, which correspond to low-dimensional ($K = 100$), mid-dimensional ($K = 200$), and high-dimensional ($K = 300$) IVs, respectively. 
The resulting empirical sizes and powers for $K=100$ are summarized in Figures \ref{figure1}-\ref{figure2}. 
Focusing on the homoskedastic setting in Figure \ref{figure1}, panels (a)-(d) reveal that all five methods exhibit stable sizes across various conditions, regardless of identification strength or sparsity. Nevertheless, while BCCH reliably controls size, it can be overly conservative, potentially leading to under-detection of significant signals. In terms of power, the Fisher-combination test emerges as the strongest performer, consistently surpassing the other approaches in Example 1 across panels (a)-(d). By leveraging information from both $\mathrm{JAR}$ and BCCH$\_$Asy, Fisher proves robust to both sparse and dense structures. 
BCCH$\_$Asy and $\mathrm{JAR}$ exhibit comparable performance overall: BCCH$\_$Asy ranks second when IVs are sparse (panels (a) and (c)), while $\mathrm{JAR}$ ranks second when IVs are dense (panels (b) and (d)). This behavior aligns with the general insight that quadratic-type tests ($\mathrm{JAR}$) are well-suited to dense signals, whereas maximum-type tests (BCCH$\_$Asy) excel at detecting sparse alternatives. 
RJAR shows the weakest power under sparse conditions, while BCCH performs worst under dense conditions. 
Moreover, $\mathrm{JAR}$ offers higher power than RJAR and employs a simpler test statistic, avoiding the need for selecting a regularization parameter. Finally, BCCH$\_$Asy enhances  power in both sparse and dense settings with its less conservative critical value compared to BCCH. 
Figure \ref{figure2} presents similar results to those in Figure \ref{figure1}, demonstrating that our proposed methods are robust to heteroskedasticity.

\begin{center}
	Figures \ref{figure1}–\ref{figure2} should be here.
\end{center}

In addition, we examine the rejection frequencies of the five methods for $K = 200$ and $K = 300$ under both homoskedastic and heteroskedastic error structures. Due to space constraints, the corresponding results  are reported in Section~\ref{sectionb} of the Supplementary Material. These findings are qualitatively consistent with those presented in Figures~\ref{figure1}–\ref{figure2} for $K = 100$.

\begin{figure}[h]
	\centering
	\includegraphics[width=0.8\textwidth]{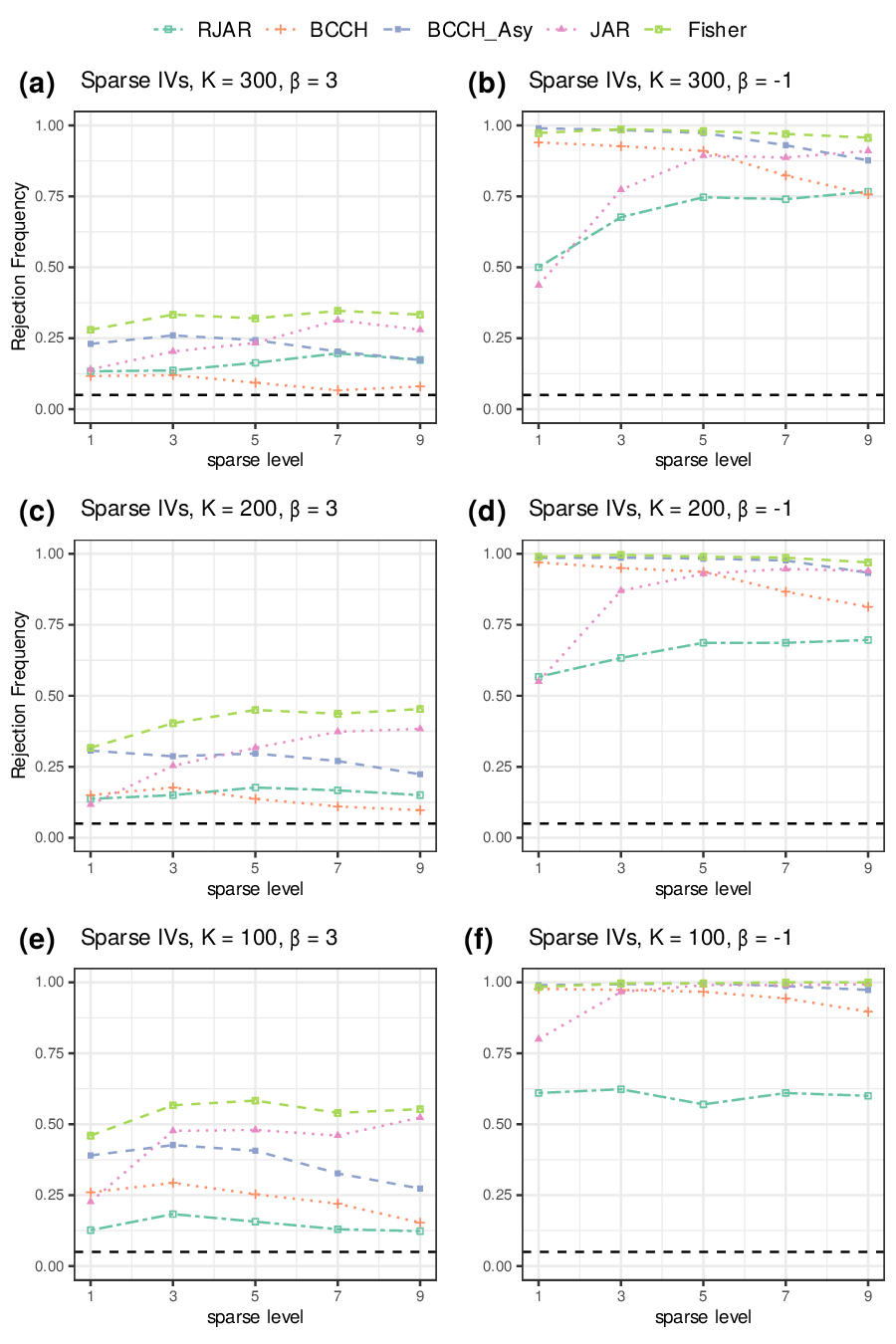}
	\caption{\setstretch{1.5}The rejection frequency curves for Example 4.1 under the sparse scenario with $\mu^2=30$ and $K \in \left\{100,200,300\right\}$: RJAR (dark green open squares, two-dash line), BCCH (orange pluses, dotted line), BCCH$\_$Asy (purple filled squares, dashed line), $\mathrm{JAR}$ (pink filled triangles, dotted line), and Fisher (light green squares, dashed line). Panels (a), (c), and (e) correspond to $\beta = 3$; panels (b), (d), and (f) correspond to $\beta = -1$. The nominal test size of 0.05 is indicated by the black horizontal line.}\label{figure7}
\end{figure}

\begin{figure}[h]
	\centering
	\includegraphics[width=0.8\textwidth]{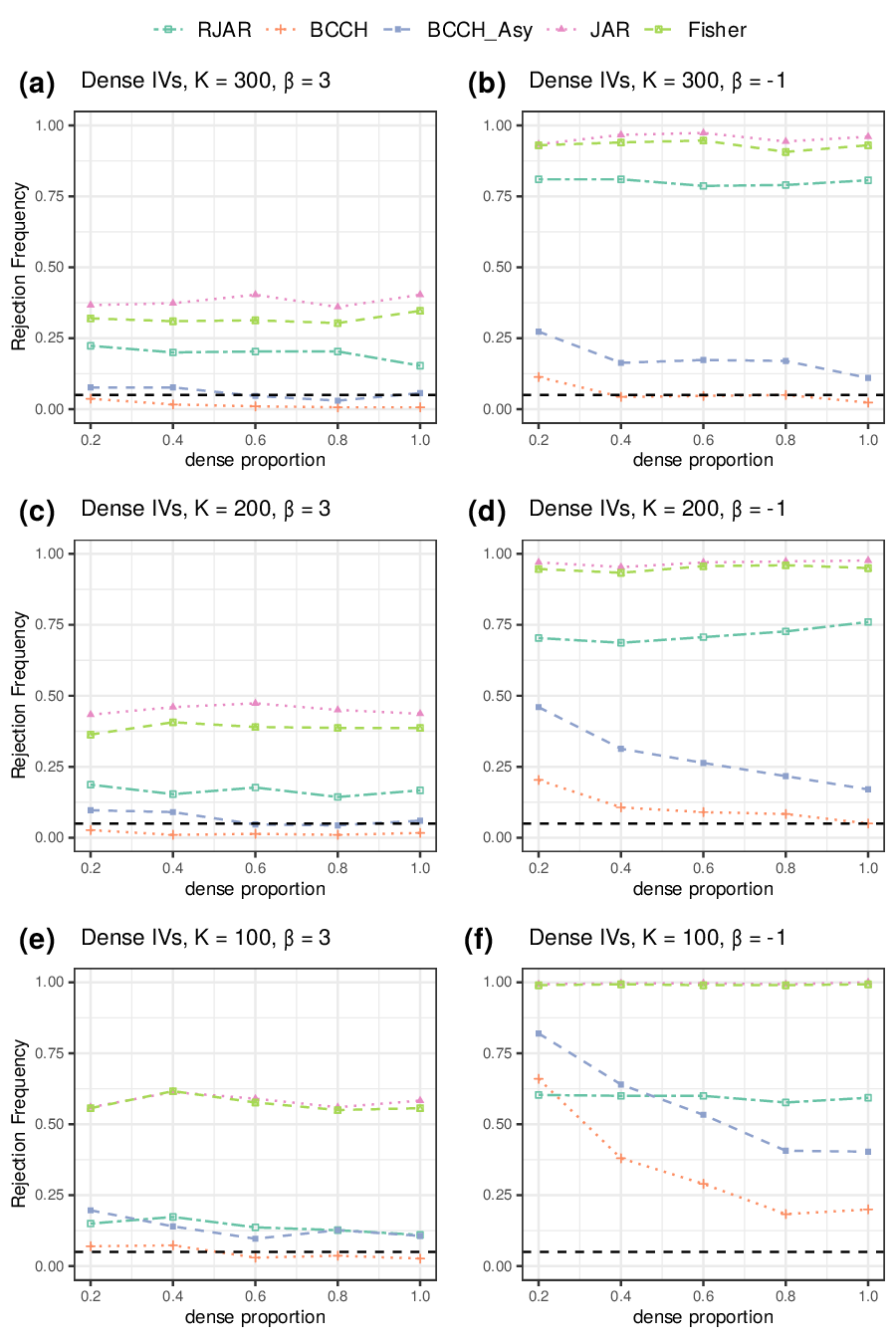}
	\caption{\setstretch{1.5}The rejection frequency curves for Example 4.2 under the dense scenario with $\mu^2=30$ and $K \in \left\{100,200,300\right\}$: RJAR (dark green open squares, two-dash line), BCCH (orange pluses, dotted line), BCCH$\_$Asy (purple filled squares, dashed line), $\mathrm{JAR}$ (pink filled triangles, dotted line), and Fisher (light green squares, dashed line). Panels (a), (c), and (e) correspond to $\beta = 3$; panels (b), (d), and (f) correspond to $\beta = -1$. The nominal test size of 0.05 is indicated by the black horizontal line.}\label{figure8}
\end{figure}

\begin{center}
	Figures \ref{figure7}–\ref{figure8} should be here.
\end{center}

To investigate how the five methods respond to varying degrees of instrumental-variable  sparsity, we conduct additional analyses using Example 4. 
Figure \ref{figure7} presents the results for a sparse IV setting under weak identification ($\mu^2 = 30$) across different number of IVs, $K \in \{100, 200, 300\}$. Panels (a), (c), and (e) show the rejection frequencies for $\beta = 3$. In all three panels, the Fisher test demonstrates the highest power. $\mathrm{JAR}$ and BCCH$\_$Asy perform similarly, while RJAR exhibits the lowest power when $K = 100$, and BCCH has the lowest power for $K = 200$ and $K = 300$. Panels (b), (d), and (f) present the results for $\beta = -1$, where Fisher test again achieves the highest power across all settings, and RJAR consistently shows the lowest power across sample sizes.
Figure \ref{figure8} summarizes the corresponding results for a dense IV setting under weak identification with $K \in \left\{100, 200, 300\right\}$. In all six panels, both the Fisher and $\mathrm{JAR}$ tests demonstrate consistently high power. The RJAR test ranks third in performance, while the BCCH test exhibits the lowest power.

\subsection{Comparison of computation time with RJAR}\label{section4.2}
In this subsection, we assess the computational efficiency of the proposed $\mathrm{JAR}$ procedure and the RJAR method \citep{dovi2024ridge} by recording their average running times across varying instrument dimensions ($K$) and sparsity levels. Following the previous setup, we set $q = 0.03K$ for the sparse scenario and $q = 0.6K$ for the dense scenario. The sample size is fixed at $n = 200$, with the concentration parameter set to $\mu^2 = 30$, $a_0 = 0$ to ensure homoskedasticity, and the number of instruments $K \in \{100, 200, 300\}$. Each configuration is replicated 100 times. Table~\ref{table3} reports the average running times of both methods across different values of $K$ under the two sparsity settings.

\begin{center}
Table~\ref{table3} should be here.
\end{center}

As shown in Table~\ref{table3}, the proposed $\mathrm{JAR}$ procedure consistently runs faster than RJAR across all tested settings. The higher computational cost of RJAR primarily attributed to its reliance on computing a ridge-regularized projection matrix and selecting the regularization parameter through an optimization procedure. In contrast, $\mathrm{JAR}$ circumvents these additional steps, leading to substantially lower computational overhead.

\subsection{Comparison of critical value with BCCH}\label{section4.3}

\begin{figure}[h]
	\centering
	\includegraphics[width=0.5\textwidth]{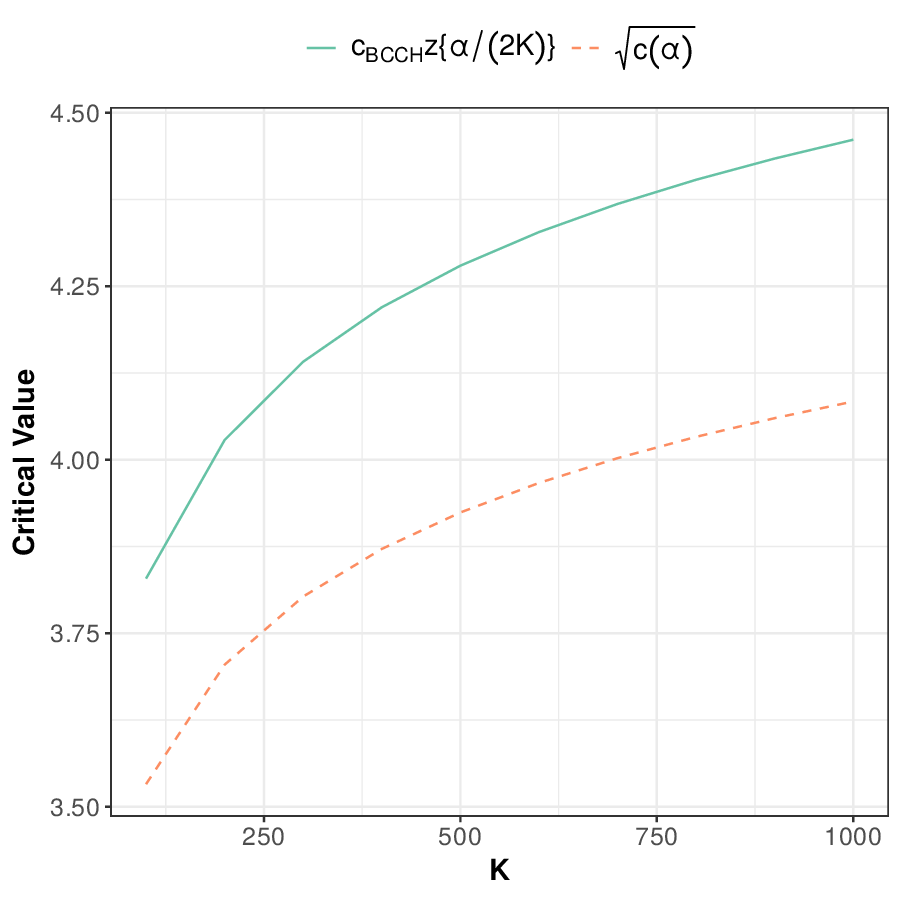}
	\caption{\setstretch{1.5} The critical value curves for \(M_n\): the BCCH-based threshold (dark green, solid line), and our refined threshold (orange, dashed line), varing the number of instruments $K$, at the significance level of 0.05.}\label{figure9}
\end{figure}

In this subsection, we compare the BCCH-based threshold proposed by \cite{belloni2012sparse} with our refined threshold for the maximum-type test statistic \(M_n\), as defined in \eqref{maximum}. 
Recall the BCCH-based threshold takes the form $c_{\mathrm{BCCH}}  z\left({ \alpha / 2K}\right)$, with $c_{\mathrm{BCCH}}=1.1$. 
And our refined critical value is given by
$\sqrt{c(\alpha)} = \left[2 (\log K) - \{\log \left(\log K \right)\}+ q_\alpha\right]^{-1/2}$, with $q_\alpha = -(\log\pi) - 2 \log \left\{\log (1 - \alpha)^{-1}\right\}$.
To facilitate a comprehensive comparison, we fix the significance level at $0.05$ and vary the number of the instrumental variables $K \in \left\{100, 200, 300, 400, 500, 600, 700, 800, 900, 1000 \right\}$. Figure~\ref{figure9} displays the curves of the two critical values of $M_n$.

\begin{center}
Figure \ref{figure9} should be here.
\end{center}

As illustrated in Figure~\ref{figure9}, both approaches yield critical values that increase with $K$. However, the BCCH-based threshold is 
consistently larger than our proposed critical value, leading to a more conservative procedure that lowers the probability of detecting true effects. In contrast, our refined critical value is less conservative, potentially providing higher power in practice.

\section{Empirical application}\label{section5}
In this section, we apply instrumental variable regression to analyze wage differentials between immigrants and natives among college-educated and high school-educated workers at the city level, following the framework of \cite{card2009immigration}. The dataset is obtained from the replication materials of \cite{goldsmith2020bartik}, and is also employed by \cite{dovi2024ridge}. We consider the following model:
\begin{equation}
	y_{is} = \beta_s X_{is} + \bm{\Gamma}_s^{\top} \bm{W}_i + \varepsilon_{is},
\end{equation}
where $y_{is}$ denotes the difference in residual log wages between immigrants and natives within skill group $s \in \left\{h, c\right\}$ (high school or college equivalent) and city $i$, for $i = 1, \dots, 124$. Residual log wages refer to log wages adjusted for education, age, gender, race, and ethnicity, as discussed in \cite{card2009immigration}.
The parameter $\beta_s$ is the coefficient of interest and can be interpreted as the negative of the inverse elasticity of substitution between immigrants and natives within skill group $s$. The vector $\bm{W}_i \in \mathbb{R}^9$ represents control variables, and $\varepsilon_{is}$ is the structural error term. The endogenous regressor $X_{is}$ is defined as the log ratio of immigrant to native hours worked in skill group $s$ in city $i$, including both men and women. As noted by \cite{card2009immigration}, unobserved city-specific factors may simultaneously affect relative wages and relative employment across skill groups. These confounding factors can bias the estimation of the inverse elasticity of substitution, motivating the use of IV methods.

We consider two IV setups in our analysis. The first follows the approach of \cite{card2009immigration}, using the $K = 38$ source countries of immigrants as instruments. The second setup, inspired by \cite{blandhol2022tsls} and \cite{dovi2024ridge}, extends the first by interacting these 38 instruments with the control variables, yielding a total of $K = 342$ instruments. 
We apply our proposed methods, the $\mathrm{JAR}$ test in Equation (\ref{quadratic}), the BCCH$\_$Asy procedure in Equation (\ref{maximum}), and the Fisher test in Equation (\ref{fisher}), alongside two benchmark procedures: RJAR \citep{dovi2024ridge} and BCCH \citep{belloni2012sparse}, to analyze the dataset. For each method, we construct a 95\% confidence set for $\beta_s$.
Specifically, we evaluate a grid of 100 candidate values $\beta_{s,0}$ and invert the tests to identify the subset of values for which the null hypothesis is not rejected at the 0.05 significance level. The resulting confidence intervals for $\beta_s$ under the two IV setups are reported in Tables \ref{table1} and \ref{table2}.

\begin{center}
	Tables \ref{table1} and \ref{table2} should be here.
\end{center}


Table \ref{table1} presents the results for the original instrument set $\bm{Z}_{38}$. For the high school group, the $\mathrm{JAR}$ and Fisher tests yield the shortest confidence intervals among the five methods, while the BCCH method produces the widest interval. In the college group, the $\mathrm{JAR}$ test again provides the shortest confidence interval, whereas the RJAR method yields the longest, even exceeding that of BCCH. This suggests that the first-stage relationship for the college group may be particularly sparse, a conclusion also noted by  \cite{dovi2024ridge}.
Table \ref{table2} reports the results for the extended instrument set $\bm{Z}_{342}$. In the high school group, the $\mathrm{JAR}$ and Fisher tests again produce the shortest confidence intervals, while BCCH results in the widest. For the college group, the $\mathrm{JAR}$ test continues to offer the shortest interval, with the Fisher test following closely. Notably, the BCCH$\_$Asy method yields a narrower interval than RJAR, while the BCCH method once again has the largest confidence interval among the five methods.

\section{Conclusions and discussions}\label{section6}
This paper investigates inference in instrumental variable regression models with high-dimensional weak  instrumental variables. 
Our proposed methods remain valid under arbitrary instrument strength and heteroskedasticity, complementing existing approaches in the many-weak-instruments literature. 
Depending on the structure of the instruments, we develop three complementary procedures. For dense IV settings, we introduce a tuning-free, computationally efficient quadratic-type test. In sparse settings, we refine the BCCH test by deriving its asymptotic null distribution and adopting a less conservative critical value to enhance power. When the sparsity of first-stage model is unknown, we propose a data-driven combination test using Fisher’s method.  

 
As with many other works, we assume the observations are independent. It would be very helpful to develop procedures for dependent observations. A recent investigation is conducted by \cite{dovi2025inference}. Further our theoretical results for $\mathrm{JAR}$ require that $K\rightarrow \infty$. Recently \cite{lim2024dimension} developed a dimension-agnostic bootstrap Anderson-Rubin test procedure. It would be interesting to investigate these important issues.

\bigskip

%
%
%
%





\bibliographystyle{apalike}
\bibliography{refe}

\begin{thebibliography}{}

\bibitem[Anatolyev and Gospodinov, 2011]{anatolyev2011specification}
Anatolyev, S. and Gospodinov, N. (2011).
\newblock Specification testing in models with many instruments.
\newblock {\em Econometric Theory}, 27(2):427--441.

\bibitem[Anderson and Rubin, 1949]{anderson1949estimation}
Anderson, T.~W. and Rubin, H. (1949).
\newblock Estimation of the parameters of a single equation in a complete
  system of stochastic equations.
\newblock {\em The Annals of Mathematical Statistics}, 20(1):46--63.

\bibitem[Andrews and Stock, 2007]{andrews2007testing}
Andrews, D.~W. and Stock, J.~H. (2007).
\newblock Testing with many weak instruments.
\newblock {\em Journal of Econometrics}, 138(1):24--46.

\bibitem[Andrews et~al., 2019]{andrews2019weak}
Andrews, I., Stock, J.~H., and Sun, L. (2019).
\newblock Weak instruments in instrumental variables regression: Theory and
  practice.
\newblock {\em Annual Review of Economics}, 11(1):727--753.

\bibitem[Bekker, 1994]{bekker1994alternative}
Bekker, P.~A. (1994).
\newblock Alternative approximations to the distributions of instrumental
  variable estimators.
\newblock {\em Econometrica: Journal of the Econometric Society},
  62(3):657--681.

\bibitem[Belloni et~al., 2012]{belloni2012sparse}
Belloni, A., Chen, D., Chernozhukov, V., and Hansen, C. (2012).
\newblock Sparse models and methods for optimal instruments with an application
  to eminent domain.
\newblock {\em Econometrica}, 80(6):2369--2429.

\bibitem[Blandhol et~al., 2022]{blandhol2022tsls}
Blandhol, C., Bonney, J., Mogstad, M., and Torgovitsky, A. (2022).
\newblock When is tsls actually late?
\newblock {\em NBER Working Paper}, (w29709).

\bibitem[Bun et~al., 2020]{bun2020finite}
Bun, M.~J., Farbmacher, H., and Poldermans, R.~W. (2020).
\newblock Finite sample properties of the gmm anderson--rubin test.
\newblock {\em Econometric Reviews}, 39(10):1042--1056.

\bibitem[Cai et~al., 2013]{cai2013two}
Cai, T., Liu, W., and Xia, Y. (2013).
\newblock Two-sample covariance matrix testing and support recovery in
  high-dimensional and sparse settings.
\newblock {\em Journal of the American Statistical Association},
  108(501):265--277.

\bibitem[Cai et~al., 2014]{tony2014two}
Cai, T., Liu, W., and Xia, Y. (2014).
\newblock Two-sample test of high dimensional means under dependence.
\newblock {\em Journal of the Royal Statistical Society Series B: Statistical
  Methodology}, 76(2):349--372.

\bibitem[Cao et~al., 2018]{cao2018two}
Cao, Y., Lin, W., and Li, H. (2018).
\newblock Two-sample tests of high-dimensional means for compositional data.
\newblock {\em Biometrika}, 105(1):115--132.

\bibitem[Card, 2009]{card2009immigration}
Card, D. (2009).
\newblock Immigration and inequality.
\newblock {\em American Economic Review}, 99(2):1--21.

\bibitem[Carrasco and Tchuente, 2016]{carrasco2016regularization}
Carrasco, M. and Tchuente, G. (2016).
\newblock Regularization based anderson rubin tests for many instruments.
\newblock {\em University of Kent, School of Economics Discussion Papers},
  (1608):1--34.

\bibitem[Chao et~al., 2014]{chao2014testing}
Chao, J.~C., Hausman, J.~A., Newey, W.~K., Swanson, N.~R., and Woutersen, T.
  (2014).
\newblock Testing overidentifying restrictions with many instruments and
  heteroskedasticity.
\newblock {\em Journal of Econometrics}, 178:15--21.

\bibitem[Chen et~al., 2024]{chen2024high}
Chen, D., Feng, L., Mykland, P.~A., and Zhang, L. (2024).
\newblock High dimensional regression coefficient test with high frequency
  data.
\newblock {\em Journal of Econometrics}, page 105812.

\bibitem[Chen et~al., 2019]{chen2019two}
Chen, S.~X., Li, J., and Zhong, P.-S. (2019).
\newblock Two-sample and anova tests for high dimensional means.
\newblock {\em Annals of Statistics}, 47(3):1443--1474.

\bibitem[Chen and Qin, 2010]{chen2010two}
Chen, S.~X. and Qin, Y.-L. (2010).
\newblock A two-sample test for high-dimensional data with applications to
  gene-set testing.
\newblock {\em Annals of Statistics}, 38:808--835.

\bibitem[Crudu et~al., 2021]{crudu2021inference}
Crudu, F., Mellace, G., and S{\'a}ndor, Z. (2021).
\newblock Inference in instrumental variable models with heteroskedasticity and
  many instruments.
\newblock {\em Econometric Theory}, 37(2):281--310.

\bibitem[Cui et~al., 2018]{10.1214/17-AOS1573}
Cui, H., Guo, W., and Zhong, W. (2018).
\newblock {Test for high-dimensional regression coefficients using refitted
  cross-validation variance estimation}.
\newblock {\em Annals of Statistics}, 46(3):958--988.

\bibitem[Davey~Smith and Hemani, 2014]{davey2014mendelian}
Davey~Smith, G. and Hemani, G. (2014).
\newblock Mendelian randomization: genetic anchors for causal inference in
  epidemiological studies.
\newblock {\em Human Molecular Genetics}, 23(R1):R89--R98.

\bibitem[Dov{\`\i}, 2025]{dovi2025inference}
Dov{\`\i}, M.-S. (2025).
\newblock Inference with high-dimensional weak instruments and the new
  keynesian phillips curve.
\newblock {\em Journal of Business \& Economic Statistics},
  (just-accepted):1--19.

\bibitem[Dov{\`\i} et~al., 2024]{dovi2024ridge}
Dov{\`\i}, M.-S., Kock, A.~B., and Mavroeidis, S. (2024).
\newblock A ridge-regularized jackknifed anderson-rubin test.
\newblock {\em Journal of Business \& Economic Statistics}, 42(3):1083--1094.

\bibitem[Fan et~al., 2025]{fan2024heteroscedasticity}
Fan, Q., Guo, Z., and Mei, Z. (2025).
\newblock A heteroscedasticity-robust overidentifying restriction test with
  high-dimensional covariates.
\newblock {\em Journal of Business \& Economic Statistics}, 43(2):413--422.

\bibitem[Feng et~al., 2024]{feng2022asymptotic}
Feng, L., Jiang, T., Li, X., and Liu, B. (2024).
\newblock Asymptotic independence of the sum and maximum of dependent random
  variables with applications to high-dimensional tests.
\newblock {\em Statistica Sinica}, 34:1745--1763.

\bibitem[Fisher, 1970]{fisher1970statistical}
Fisher, R.~A. (1970).
\newblock {\em Statistical Methods for Research Workers}.
\newblock Springer.

\bibitem[Fuller, 1977]{fuller1977some}
Fuller, W.~A. (1977).
\newblock Some properties of a modification of the limited information
  estimator.
\newblock {\em Econometrica}, 45(4):939--953.

\bibitem[Glymour, 2006]{glymour2006natural}
Glymour, M.~M. (2006).
\newblock Natural experiments and instrumental variable analyses in social
  epidemiology.
\newblock {\em Methods in Social Epidemiology}, 1:429.

\bibitem[Gold et~al., 2020]{gold2020inference}
Gold, D., Lederer, J., and Tao, J. (2020).
\newblock Inference for high-dimensional instrumental variables regression.
\newblock {\em Journal of Econometrics}, 217(1):79--111.

\bibitem[Goldsmith-Pinkham et~al., 2020]{goldsmith2020bartik}
Goldsmith-Pinkham, P., Sorkin, I., and Swift, H. (2020).
\newblock Bartik instruments: What, when, why, and how.
\newblock {\em American Economic Review}, 110(8):2586--2624.

\bibitem[Guo and Chen, 2016]{guo2016tests}
Guo, B. and Chen, S.~X. (2016).
\newblock Tests for high dimensional generalized linear models.
\newblock {\em Journal of the Royal Statistical Society Series B: Statistical
  Methodology}, 78(5):1079--1102.

\bibitem[Guo et~al., 2022]{guo2022conditional}
Guo, W., Zhong, W., Duan, S., and Cui, H. (2022).
\newblock Conditional test for ultrahigh dimensional linear regression
  coefficients.
\newblock {\em Statistica Sinica}, 32(3):1381--1409.

\bibitem[Hahn and Hausman, 2003]{hahn2003weak}
Hahn, J. and Hausman, J. (2003).
\newblock Weak instruments: Diagnosis and cures in empirical econometrics.
\newblock {\em American Economic Review}, 93(2):118--125.

\bibitem[Hausman et~al., 2012]{hausman2012instrumental}
Hausman, J.~A., Newey, W.~K., Woutersen, T., Chao, J.~C., and Swanson, N.~R.
  (2012).
\newblock Instrumental variable estimation with heteroskedasticity and many
  instruments.
\newblock {\em Quantitative Economics}, 3(2):211--255.

\bibitem[Kleibergen, 2002]{kleibergen2002pivotal}
Kleibergen, F. (2002).
\newblock Pivotal statistics for testing structural parameters in instrumental
  variables regression.
\newblock {\em Econometrica}, 70(5):1781--1803.

\bibitem[Lawlor et~al., 2008]{lawlor2008mendelian}
Lawlor, D.~A., Harbord, R.~M., Sterne, J.~A., Timpson, N., and Davey~Smith, G.
  (2008).
\newblock Mendelian randomization: using genes as instruments for making causal
  inferences in epidemiology.
\newblock {\em Statistics in Medicine}, 27(8):1133--1163.

\bibitem[Li et~al., 2025]{li2024power}
Li, D., Xue, L., Yang, H., and Yu, X. (2025).
\newblock Power-enhanced two-sample mean tests for high-dimensional microbiome
  compositional data.
\newblock {\em Biometrics}, 81(2):ujaf034.

\bibitem[Li and Chen, 2012]{10.1214/12-AOS993}
Li, J. and Chen, S.~X. (2012).
\newblock {Two sample tests for high-dimensional covariance matrices}.
\newblock {\em Annals of Statistics}, 40(2):908 -- 940.

\bibitem[Lim et~al., 2024]{lim2024dimension}
Lim, D., Wang, W., and Zhang, Y. (2024).
\newblock A dimension-agnostic bootstrap anderson-rubin test for instrumental
  variable regressions.
\newblock {\em arXiv preprint arXiv:2412.01603}.

\bibitem[Littell and Folks, 1971]{littell1971asymptotic}
Littell, R.~C. and Folks, J.~L. (1971).
\newblock Asymptotic optimality of fisher's method of combining independent
  tests.
\newblock {\em Journal of the American Statistical Association},
  66(336):802--806.

\bibitem[Littell and Folks, 1973]{littell1973asymptotic}
Littell, R.~C. and Folks, J.~L. (1973).
\newblock Asymptotic optimality of fisher's method of combining independent
  tests ii.
\newblock {\em Journal of the American Statistical Association},
  68(341):193--194.

\bibitem[Ma et~al., 2021]{ma2021global}
Ma, R., Cai, T., and Li, H. (2021).
\newblock Global and simultaneous hypothesis testing for high-dimensional
  logistic regression models.
\newblock {\em Journal of the American Statistical Association},
  116(534):984--998.

\bibitem[Mikusheva and Sun, 2022]{mikusheva2022inference}
Mikusheva, A. and Sun, L. (2022).
\newblock Inference with many weak instruments.
\newblock {\em The Review of Economic Studies}, 89(5):2663--2686.

\bibitem[Moreira, 2003]{moreira2003conditional}
Moreira, M.~J. (2003).
\newblock A conditional likelihood ratio test for structural models.
\newblock {\em Econometrica}, 71(4):1027--1048.

\bibitem[Moreira, 2009]{moreira2009tests}
Moreira, M.~J. (2009).
\newblock Tests with correct size when instruments can be arbitrarily weak.
\newblock {\em Journal of Econometrics}, 152(2):131--140.

\bibitem[Narum, 2006]{narum2006beyond}
Narum, S.~R. (2006).
\newblock Beyond bonferroni: less conservative analyses for conservation
  genetics.
\newblock {\em Conservation Genetics}, 7:783--787.

\bibitem[Newey and Windmeijer, 2009]{newey2009generalized}
Newey, W.~K. and Windmeijer, F. (2009).
\newblock Generalized method of moments with many weak moment conditions.
\newblock {\em Econometrica}, 77(3):687--719.

\bibitem[Papaspiliopoulos, 2020]{papaspiliopoulos2020high}
Papaspiliopoulos, O. (2020).
\newblock {\em High-Dimensional Probability: An Introduction with Applications
  in Data Science}.
\newblock Taylor \& Francis.

\bibitem[Phillips and Gao, 2017]{phillips2017structural}
Phillips, P.~C. and Gao, W.~Y. (2017).
\newblock Structural inference from reduced forms with many instruments.
\newblock {\em Journal of Econometrics}, 199(2):96--116.

\bibitem[Srivastava et~al., 2013]{srivastava2013two}
Srivastava, M.~S., Katayama, S., and Kano, Y. (2013).
\newblock A two sample test in high dimensional data.
\newblock {\em Journal of Multivariate Analysis}, 114:349--358.

\bibitem[Wang and Zivot, 1998]{wang1998inference}
Wang, J. and Zivot, E. (1998).
\newblock Inference on structural parameters in instrumental variables
  regression with weak instruments.
\newblock {\em Econometrica}, 66(6):1389--1404.

\bibitem[Xu et~al., 2016]{xu2016adaptive}
Xu, G., Lin, L., Wei, P., and Pan, W. (2016).
\newblock An adaptive two-sample test for high-dimensional means.
\newblock {\em Biometrika}, 103(3):609--624.

\bibitem[Yu et~al., 2024]{yu2024fisher}
Yu, X., Li, D., and Xue, L. (2024).
\newblock Fisher’s combined probability test for high-dimensional covariance
  matrices.
\newblock {\em Journal of the American Statistical Association},
  119(545):511--524.

\bibitem[Zhang and Cheng, 2017]{zhang2017simultaneous}
Zhang, X. and Cheng, G. (2017).
\newblock Simultaneous inference for high-dimensional linear models.
\newblock {\em Journal of the American Statistical Association},
  112(518):757--768.

\bibitem[Zhong and Chen, 2011]{zhong2011tests}
Zhong, P.-S. and Chen, S.~X. (2011).
\newblock Tests for high-dimensional regression coefficients with factorial
  designs.
\newblock {\em Journal of the American Statistical Association},
  106(493):260--274.

\bibitem[Zhu, 2018]{zhu2018sparse}
Zhu, Y. (2018).
\newblock Sparse linear models and l1-regularized 2sls with high-dimensional
  endogenous regressors and instruments.
\newblock {\em Journal of Econometrics}, 202(2):196--213.

\end{thebibliography}


\newpage 

\section*{Tables and figures}


\begin{table}[h]
\centering
\renewcommand{\arraystretch}{1.3}
\setlength{\tabcolsep}{18pt} 
\caption{The average computation time (Unit: second)  of RJAR and {JAR}.}\label{table3}
\begin{tabular}{c c c c c}
\toprule
$K$ & \text{IV Structure} & \text{RJAR} & \text{JAR}  \\
\midrule
\multirow{2}{*}{100} & Sparse &13.9842&0.0379 \\
                     & Dense  &13.9971&0.0430\\
\multirow{2}{*}{200} & Sparse &17.5740&0.0649 \\
                     & Dense  &17.6034&0.0688 \\
\multirow{2}{*}{300} & Sparse &24.6710&0.0985 \\
                     & Dense  & 24.6944&0.0929 \\
\bottomrule
\end{tabular}
\end{table}

\begin{table}[h]
	\centering
	\renewcommand{\arraystretch}{1}
	\setlength{\tabcolsep}{4pt} 
	\caption{95\% confidence intervals (CI) with $\bm{Z}_{38}$ as instruments.}\label{table1}
	\footnotesize 
	\begin{tabular}{lccccc}
		\toprule
		& RJAR & BCCH & BCCH$\_$Asy & $\mathrm{JAR}$ & Fisher \\
		\midrule
		\multicolumn{6}{c}{High-School Workers} \\
		\midrule
		CI & \textbf{[-0.0803, -0.0157]} & \textbf{[-0.0722, -0.0035]} & \textbf{[-0.0641, -0.0076]} & \textbf{[-0.0561, -0.0197]} & \textbf{[-0.0561, -0.0197]} \\
		Length & 0.0646 & 0.0687 & 0.0565 & 0.0364 & 0.0364 \\
		\midrule
		\multicolumn{6}{c}{College Workers} \\
		\midrule
		CI & \textbf{[-0.1369, 0.0086]} & \textbf{[-0.1288, -0.0157]} & \textbf{[-0.1167, -0.0237]} & \textbf{[-0.1005, -0.0480]} & \textbf{[-0.1005, -0.0439]} \\
		Length & 0.1455 & 0.1131 & 0.0929 & 0.0525 & 0.0566 \\
		\bottomrule
	\end{tabular}
\end{table}

\begin{table}[h]
	\centering
	\renewcommand{\arraystretch}{1}
	\setlength{\tabcolsep}{4pt} 
	\caption{95\% confidence intervals (CI) with $\bm{Z}_{342}$ as instruments.}\label{table2}
	\footnotesize 
	\begin{tabular}{lccccc}
		\toprule
		& RJAR & BCCH & BCCH$\_$Asy & $\mathrm{JAR}$ & Fisher \\
		\midrule
		\multicolumn{6}{c}{High-School Workers} \\
		\midrule
		CI & \textbf{[-0.0763, -0.0116]} & \textbf{[-0.0843, 0.0045]} & \textbf{[-0.0722, -0.0035]} & \textbf{[-0.0561, -0.0197]} & \textbf{[-0.0561, -0.0197]} \\
		Length & 0.0647& 0.0888& 0.0687& 0.0364& 0.0364 \\
		\midrule
		\multicolumn{6}{c}{College Workers} \\
		\midrule
		CI & \textbf{[-0.1086, 0.0086]} & \textbf{[-0.1490, -0.0035]} & \textbf{[-0.1288, -0.0157]} & \textbf{[-0.1005, -0.0439]} & \textbf{[-0.1045, -0.0399]} \\
		Length & 0.1172& 0.1455& 0.1131& 0.0566& 0.0646 \\
		\bottomrule
	\end{tabular}
\end{table}

\end{document}